\documentclass[fleqn,usenatbib]{mnras}

\usepackage[T1]{fontenc}
\usepackage{ae,aecompl}

\usepackage{hyperref}
\usepackage{booktabs}
\usepackage{multirow}
\usepackage{pgf}
\usepackage{subcaption}
\usepackage{threeparttable}
\RequirePackage{fix-cm}

\usepackage{graphicx}	
\usepackage{amsmath}	
\usepackage{amssymb}	

\usepackage{newtxtext,newtxmath}

\newcommand{\angs}{\textup{\AA}}
\newcommand{\DPA}{${\Delta\mbox{PA}}$}
\newcommand{\Ha}{\ion{H}{$\;\!\!\alpha$}}
\newcommand{\Hb}{\ion{H}{$\;\!\!\beta$}}

\newcommand{\Sii}{[\ion{S}{ii}]}
\newcommand{\Reff}{$R_{\mathrm e}$}

\title[Fuelling galactic centres with misaligned gas]{Fuelling the central region of galaxies with misaligned gas accretion}

\author[J.\ L.\ Tous et al.]{
J.\ L.\ Tous$^{1}$\thanks{E-mail: j.l.tous-mayol@soton.ac.uk}, S. I. Raimundo$^1$, R.\ Riffel$^{2}$, A. Puglisi$^1$, F. Shankar$^1$, M. Vestergaard$^{3, 4}$\\
$^{1}$School of Physics and Astronomy, University of Southampton, Highfield, Southampton SO17 1BJ, UK\\
$^{2}$Departamento de Astronomia, Instituto de F\'\i sica, Universidade Federal do Rio Grande do Sul, CP 15051, 91501-970, Porto Alegre, RS, Brazil\\
$^{3}$DARK, Niels Bohr Institute, University of Copenhagen, Jagtvej 155, Copenhagen N, DK-2200, Denmark\\
$^{4}$Steward Observatory, University of Arizona, 933 N Cherry Ave, Tucson, AZ 85721, USA
}
\date{Accepted XXX. Received YYY; in original form ZZZ}

\pubyear{2026}

\begin{document}
\label{firstpage}
\pagerange{\pageref{firstpage}--\pageref{lastpage}}
\maketitle

\begin{abstract}
Recent studies have shown that the accretion of kinematically misaligned gas fuels the central reservoir of galaxies, triggering nuclear activity and star formation. In this work, we show that this specific accretion channel can only sustain main‑sequence levels of star formation in galaxies with $M_* \lesssim 10^{10}\;{\rm M_\odot}$. Using a sample of $201$ kinematically misaligned galaxies, and a comparison sample of $3689$ aligned galaxies, from the Mapping Nearby Galaxies at Apache Point Observatory survey, we investigate the impact of kinematically misaligned gas on star formation across stellar mass. We characterise the global specific star formation rate, central concentration and mass of ionized gas of our samples, and derive radial profiles of specific star formation rate and ionized gas mass surface density. We find that misaligned galaxies exhibit more centrally concentrated ionized gas than their aligned counterparts, with the strongest enhancement occurring at intermediate masses ($10^{10}$–$10^{10.6}\;{\rm M_\odot}$), where the central ionized‑gas density peaks. In galaxies less massive than $10^{10}\;{\rm M_\odot}$, misaligned gas fuels nuclear star formation at rates typical of star-forming systems. At higher masses, however, the impact of this newly accreted gas is diluted by the larger pre-existing stellar mass in the central regions, limiting its ability to rejuvenate the star formation activity in these galaxies. Our results also show that misaligned galaxies with or without nuclear activity exhibit similar central concentrations of ionized gas, suggesting that most have a reservoir capable of fuelling their supermassive black holes over timescales longer than typical nuclear activity episodes.
\end{abstract}

\begin{keywords}
galaxies: star formation; galaxies: evolution; galaxies: active; galaxies: kinematics and dynamics; galaxies: statistics
\end{keywords}

\section{Introduction}
\label{S:intro}

With just a small fraction of the baryonic content of the Universe locked up in stars \citep{Fukugita2004, Shull2012}, a substantial reservoir of baryons remains in the form of gas, potentially available to fuel star formation in local galaxies. However, not all of this gas is readily accessible, as it exists in different phases and reservoirs. The interstellar medium comprises cold molecular gas, warm atomic gas, and hot ionised gas, each playing a distinct role in the star formation cycle. While cold molecular gas is the immediate fuel for star formation, the transition from warmer or ionised phases into the molecular phase depends on complex cooling processes, environmental conditions, and dynamical triggers \citep{Glover2012, Krumholz2012, Feldmann2020, Hunt2020, Girichidis2021}. Moreover, a significant fraction of baryons resides in the circumgalactic and intergalactic medium \citep[e.g.,][]{Solanes2001, Haider2016, Tumlinson2017}, where they may eventually accrete onto galaxies through different processes, but only after undergoing substantial cooling and angular momentum redistribution \citep{Keres2005}. Thus, investigating how gas is accreted, transformed into stars, and expelled from galaxies---and transitions between these phases in the process---is crucial to understanding the mechanisms that drive star formation quenching and rejuvenation.

It is now well established that the star formation rate (SFR) and the stellar mass ($M_*$) of star-forming galaxies follow a tight correlation known as the main sequence of star-forming galaxies. This relation is observed across a wide range of redshifts \citep{Whitaker2012}. As they grow in mass, star-forming galaxies evolve along the main sequence, with their SFRs increasing almost linearly with stellar mass. Multiple quenching mechanisms can reduce the SFR of galaxies, make them leave the main sequence, and eventually become quiescent. Supernova feedback and stellar winds are typically invoked to regulate star formation in low-mass galaxies, while at high mass, AGN feedback and halo mass are thought to be the main drivers of galaxy quenching \citep{Dekel2006, Shankar2006, Zolotov2015, Tachella2016, Fu2025}. In addition, satellite galaxies are especially vulnerable to environmental quenching through mechanisms such as ram pressure stripping \citep{Gunn&Gott1972} or strangulation \citep{Larson1980}. However, if a quenching galaxy accretes fresh gas, e.g. by interacting with a gas-rich companion, its SFR can increase again, resulting in a rejuvenation phase in which it moves back toward the main sequence.

In the local Universe, star-forming galaxies tend to exhibit late-type morphologies, whereas quiescent galaxies preferentially display early-type morphologies \citep{Strateva2001}. The same processes that quench star formation in galaxies may also drive a morphological transformation, and vice versa. On one hand, without cold gas to fuel star formation in the disc and dissipate energy, galaxies are more likely to evolve toward dynamically hotter, pressure-supported systems with no spiral arms \citep{Sellwood2022}. On the other hand, the growth of a stellar spheroid stabilizes the system against gas fragmentation, hence quenching star formation \citep{Martig2009}. The link between galaxy quenching and morphology is supported, for example, by observations such as the well known morphology-density relation \citep{Dressler1980} along with the increase in the fraction of passive galaxies with increasing environmental density \citep[e.g.,][]{Tous2020, JimP22, Lopes2024, Cleland2025, Gort2025}. However, recent studies show that star formation is not a rare phenomenon in ETGs, especially among S0 galaxies \citep{Kaviraj2007, Salim2012, Mendez-Abreu2019, Tous2020, JimP22, Tous2025}. Accretion events, such as minor mergers, are an attractive solution to explain both extended star formation activity in ETGs located in the field or small groups \citep[e.g.,][]{mckelvie2018, Deeley2020, Lacerna2020, Coccato2022, Maschmann2022, Tous2024}, and localized star formation within inner rings of S0 galaxies \citep{Tous2023}. Although these events can trigger rejuvenation, galaxies rarely recover their previous morphology, as most of them promote bulge growth and inhibit the formation of spiral arms \citep{Hopkins2010, Osman2017, Zhou2024}.

Perhaps the strongest piece of evidence supporting an external origin of the star-forming gas in ETGs is that, in these systems, this component is often found decoupled from the motion of the stars. For example, \citet{Davis2011} found that $41$ per cent of the ETGs with ionized gas from the ATLAS$^{\rm 3D}$ survey \citep{Capellari2011} show a misalignment of more than $30^\circ$ between the kinematic position angles (PAs) of the gas and the stars. More recently, \citet{Rathore2022} found that in a sample of $120$ low-mass, star-forming S0 galaxies from the Mapping Nearby Galaxies at Apache Point Observatory (MaNGA) survey \citep{Bundy2015}, more than half of the galaxies present misaligned or unsettled kinematics compared to the stars. The structural similarity of these galaxies to quiescent S0s, led the authors to conclude that star formation has been rejuvenated in these galaxies rather than being the aftermath of the transition from late- to early-type morphologies. In contrast, these kinematic features are rare among late-type galaxies \citep[e.g.,][]{Bryant2019, Ristea2022, Raimundo2023}. Moreover, the observed incidence of kinematic misalignments associated with morphology is consistently reproduced across different cosmological hydrodynamical simulations \citep{Khim2020, Casanueva2022}. Although powerful outflows may affect the gas kinematics of some galaxies, especially in the central kiloparsec \citep[e.g.,][]{Ilha2019}, it is more likely that most of the galaxy-wide-scale gas misalignments arise from accretion events \citep{Raimundo2023, Winiarska2025}.

Hydrodynamical simulations have also demonstrated that mechanisms such as shocks, stellar torques, and dynamical friction can lead gas in misaligned structures to lose angular momentum and migrate toward the central regions of galaxies \citep{Hernquist1995, Thakar1996, Voort2015, Capelo2017, Blumenthal2018, Taylor2018, Khoperskov2021}. These processes have been proposed to account for the observed enhancement of central star formation in kinematically misaligned galaxies \citep{Chen2016, Xu2022}. Further evidence of the effect of kinematic misalignment on the central regions of galaxies comes from studies focussed on supermassive black holes (SMBHs). \citet{Raimundo2023} reported a statistically significant excess of active galactic nuclei (AGN) in misaligned galaxies compared to their aligned counterparts, based on a sample of $1310$ galaxies from the Sydney-AAO Multi-object Integral field spectrograph survey \citep{SAMI}. This finding established, for the first time, a direct observational link between gas misalignment and SMBH activity. Subsequently, \citet{Raimundo2025} showed that the elevated AGN fraction in misaligned systems persists even when galaxies exhibit signs of recent external interactions, hence highlighting the role of misaligned gas structures in the fuelling of SMBHs.

Although the connection between gas misalignment and the inflow of gas towards the inner regions of galaxies is strongly supported by simulations and observations, a direct correlation with SMBH growth seems less clear. On one hand, according to predictions from cosmological hydrodynamical simulations, gas misalignment can persist for $\sim 10^8-10^9$ yr \citep{Voort2015, Baker2025}. Conversely, AGN episodes are typically short-lived, with duty cycles of $\sim 10^5 - 10^8$ yr \citep{Gati2015, Schawinski2015}. This mismatch in time-scales means that while misaligned gas may provide the necessary conditions for black hole fuelling, the AGN phase may occur intermittently and be missed in a snapshot observation \citep{Duckworth2020}. On the other hand, \citet{Winiarska2025} found that AGN luminosities in misaligned galaxies are no different from those of aligned galaxies. This result suggests that AGN luminosity is driven by other factors such as the availability of fresh gas near the SMBH.

Rejuvenation events typically contribute a small fraction to a galaxy’s total stellar mass, but their impact on the present-day cosmic SFR density is much more significant than naively expected \citep{Chauke2019, Tanaka2024}. Besides, the same processes that rejuvenate galaxies, may also trigger episodes of nuclear activity and black hole growth \citep{Mallmann2018, Martin-Navarro2022, Riffel2022, Riffel2023, Rembold2024, Choi2024, Gatto2025}. Understanding these processes is therefore essential for building a comprehensive picture of galaxy evolution. In this work, we investigate how kinematically misaligned gas accretion affects the star-forming properties of nearby galaxies, with the goal of constraining its role in galaxy rejuvenation and assessing whether this mechanism operates similarly across different stellar masses. The sample of MaNGA galaxies analysed is described in Section~\ref{S:sample} along with the selection method of kinematically misaligned systems. In Section~\ref{S:results}, our results on global properties of kinematically misaligned galaxies are presented (Sections~\ref{SS:mstar_ssfr} and \ref{SS:ionized_gas}) and radial profiles are derived and analysed (Section~\ref{SS:radial_profiles}). In Section~\ref{S:discussion}, we discuss the results and present a summary of the main findings and our conclusions in Section~\ref{S:conclusion}.

The values of all cosmology-dependent quantities used in this work have been calculated by assuming a concordant flat $\Lambda$CDM universe with parameters $H_0 = 70\; {\rm km\;s^{-1}\;Mpc^{-1}}$ and $\Omega_m = 1 - \Omega_\Lambda = 0.3$.

\section{The sample}
\label{S:sample}
In this work, we use spatially resolved optical spectra from the latest data release of MaNGA  \citep{Abdurro2022}, which is part of the Sloan Digital Sky Survey IV \citep[SDSS;][]{Blanton2017}. This survey mapped the spectra of over $10 \,000$ nearby ($z < 0.15$) galaxies at least up to 1.5 times the effective radius (\Reff), covering the wavelength range of $3500-10\,000$\ \angs\, with a spectral resolution of $R \sim 2000$ \citep{Smee13}, and a spatial resolution of about one kpc at the median redshift of the sample ($z \sim 0.03$). Alongside the spectral data cubes, spatially resolved data products were generated for each galaxy \citep{Westfall2019}, from which we take the normalized elliptical radial coordinate of each galaxy's spaxels, $R/$\Reff. Maps of stellar and gas properties, and SFR are obtained from the MEGACUBES\footnote{\url{https://manga.linea.org.br/}} data sets produced by \citet{Riffel2023}, who performed a full spectral fitting stellar population synthesis on the MaNGA data cubes and fitted the profiles of the most commonly observed emission lines.

Following \citet{Raimundo2023}, we determine the kinematic misalignment angle (\DPA) for each galaxy from the difference between the PA of the stellar and gas velocity fields of each galaxy: \DPA $\;= |{\rm PA_{stars}} - {\rm PA_{gas}}|$. To do so, we use maps of gas velocity (derived from the \Ha\ emission) and stellar velocity, from which we determine the PAs by identifying the axis that maximises the velocity gradient in each map (see \citealt{Raimundo2023} for more details). For this calculation, we only use spaxels with uncertainties in the velocities lower than $30$ km/s. The values of \DPA\ range from perfect alignment at $0^\circ$, to counter-rotation at $180^\circ$ and, in this work, we consider that \DPA $\;\geq 45^\circ$ correspond to high kinematic misalignments, and \DPA  $\;< 45^\circ$ to low kinematic misalignments. Since we are interested in studying the effect of the misaligned gas in the evolution of galaxies, our sample contains all MaNGA galaxies for which it is possible to reliably measure this angle with an uncertainty lower than $30^\circ$. We obtain the global stellar mass and SFR of these galaxies from \citet{Salim2018}, who derived these properties by fitting the observed spectral energy distributions of the objects across the UV, optical, and IR domains. Our final sample contains $3890$ galaxies, of which $201$ and $3689$ exhibit, respectively, high and low kinematic misalignment\footnote{We note that $3$ galaxies from the misaligned subsample, and $37$ from the aligned subsample present an offset between the $R/$\Reff\ maps from MaNGA and the MEGACUBES. Since they represent a small fraction of the subsamples we simply exclude them from the analysis conducted in sections~\ref{SS:ionized_gas}, \ref{SS:radial_profiles}, and \ref{S:discussion}.}.

\section{Results}
\label{S:results}

In this section, we present the results from the analysis of both the global and radial distribution of star formation activity of kinematically misaligned galaxies across stellar mass.

\subsection{Global stellar mass and specific star formation rate}
\label{SS:mstar_ssfr}

The stellar mass is widely regarded as a fundamental parameter in galaxy evolution, regulating other key properties such as the star formation and metallicity. Previous studies have shown that the fraction of galaxies exhibiting strong kinematic misalignments varies systematically with both stellar mass and SFR per unit mass (specific SFR: SSFR), suggesting that long-lived kinematic misalignments are linked to specific intrinsic physical conditions \citep{Davis2011, Jin2016, Zhou2022}. 

\begin{figure*}
    \centering
	\includegraphics[width=\textwidth]{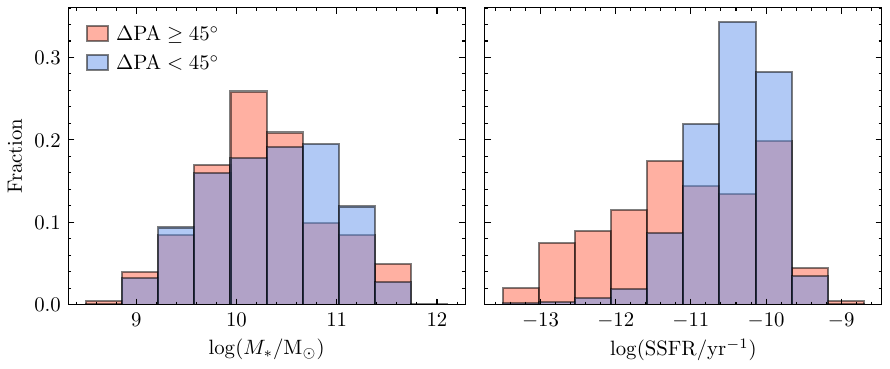}
    \caption{Distribution of global stellar mass (left) and specific star formation rate (right) of galaxies with high (red) and low (blue) kinematic misalignment. In both panels, the histograms show, for each subset, the fraction of galaxies in a given bin so each distribution is normalized to unity. For both properties, the observed differences are statistically significant according to a two-sample KS test, with $p$-values $\lesssim 0.01$.}
    \label{fig:ssfr_mass_hist}
\end{figure*}

In Fig.~\ref{fig:ssfr_mass_hist}, we show the distributions of stellar mass and SSFR for the galaxies in our sample with high kinematic misalignment in red, compared to those with low misalignment in blue. The stellar mass distribution of misaligned galaxies peaks at lower masses, is narrower, and exhibits a more pronounced positive skew than that of galaxies with low misalignment. The fraction of galaxies in both subsets increases with stellar mass, but diverges at higher masses: for highly misaligned galaxies, the fraction starts to decline above $\approx 10^{10.1}\;M_\odot$, whereas for low misalignment galaxies, it continues to rise more gradually until $\approx 10^{11}\;M_\odot$, after which it decreases. The SSFR distribution of highly misaligned galaxies is slightly bimodal, with peaks at $\approx 10^{-11.3}\;{\rm yr^{-1}}$ and $\approx 10^{-9.9}\;{\rm yr^{-1}}$, indicating the presence of both quiescent and actively star-forming populations. In contrast, galaxies with low kinematic misalignment show a narrower SSFR distribution centred around $10^{-10.4}\;{\rm yr^{-1}}$. To test the statistical significance of the observed differences, we apply a two-sample Kolmogorov–Smirnov (KS) test. The resulting $p$-value represents the probability of observing these differences under the null hypothesis---that both samples originate from the same parent population. Throughout this paper, we adopt $p \leq 0.01$ as the threshold for significance, noting that smaller values indicate stronger evidence against the null hypothesis. For both stellar mass and SSFR, the differences between the two subsets are significant, with $p$-values of $\approx 0.01$ and $\ll 0.001$, respectively.

\begin{figure}
    \centering
	\includegraphics[width=\columnwidth]{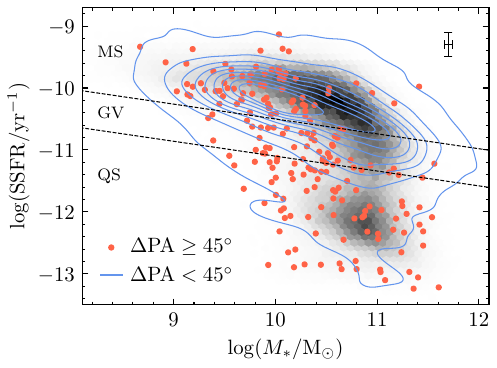}
    \caption{Distribution of galaxies with high (red dots) and low (blue contours) kinematic misalignment in the global specific star formation rate vs stellar mass plane. The contour lines are derived from a kernel density estimate of the distribution of aligned galaxies. For reference, the grey-shaded background shows the distribution of SDSS galaxies (at $z < 0.15$) binned into hexagonal cells, where darker tones indicate higher number densities. Two dashed lines split the sample into main sequence (MS), green valley (green valley), and quiescent (QS) galaxies, as specified in section~\ref{SS:mstar_ssfr}. The error bars in the top-right corner show the medians of the errors of all the galaxies in both axes.}
    \label{fig:ssfr_mass}
\end{figure}

In Fig.~\ref{fig:ssfr_mass}, we present the 2D distributions of the two galaxy subsets obtained by showing the global SSFR values as a function of the stellar mass \citep{Salim2018}. Galaxies with high kinematic misalignment are marked with red dots, while the distribution of galaxies with low misalignment is represented by blue contours derived from a kernel density estimate \citep{Silverman2018}. For reference, the background hexagonal bins in varying shades of gray indicate the distribution of galaxies from the NASA-Sloan Atlas \citep[NSA,][]{Blanton2011} catalogue from which the MaNGA sample is drawn. Two dashed lines divide the parameter space into three regimes---main sequence, green valley, and quiescent---based on their offset from the star-forming main sequence defined by \citet{Renzini2015}: $\log({\rm SSFR_{MS}}) = -0.24 \log(M_*) - 7.6$. Following \citet{Tous2024}, galaxies are assigned to the main sequence, green valley, or quiescent regimes if $\log({\rm SSFR_{MS}})$ is, respectively, $> -0.5$, $[-0.5, -1.1)$, or $\leq -1.1$. This exercise reveals that the SSFR of galaxies with high kinematic misalignment declines more steeply with stellar mass than that of galaxies with low kinematic misalignment. As a result, while the latter predominantly occupy the main sequence across the full mass range, the former transition from being concentrated in the main sequence at low masses to largely avoiding it above $M_* \gtrsim 10^{10.6}\;M_\odot$.

\begin{figure}
    \centering
	\includegraphics[width=\columnwidth]{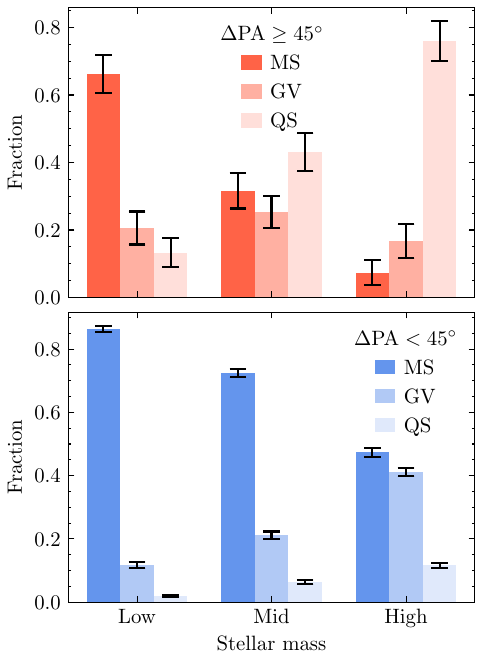}
    \caption{Fraction of galaxies with high (top) and low (bottom) kinematic misalignment in three bins of stellar mass: $M_* \leq 10^{10}\;{\rm M_\odot}$ (low), $ 10^{10}\;{\rm M_\odot} < M_* \leq 10^{10.6}\;{\rm M_\odot}$ (mid), and $M_* > 10^{10.6}\;{\rm M_\odot}$ (high). In both panels, the different tones, from darker to lighter, correspond to galaxies in the main sequence (MS), green valley (green valley), and quiescent (QS) regimes. The error bars show the $1 \sigma$ uncertainty from $1000$ bootstrap resamples with replacement of the original distributions in each bin.}
    \label{fig:fraction_vs_mass}
\end{figure}

To quantify this trend, Fig.~\ref{fig:fraction_vs_mass} shows the fraction of galaxies with high and low kinematic misalignment in our sample that belong to the main sequence, green valley, and quiescent regimes (represented by progressively lighter tones) in three bins of stellar mass: low ($M_* \leq 10^{10}\;{\rm M_\odot}$), intermediate ($ 10^{10}\;{\rm M_\odot} < M_* \leq 10^{10.6}\;{\rm M_\odot}$), and high ($M_* > 10^{10.6}\;{\rm M_\odot}$) stellar mass. At low mass, about $65$ per cent of misaligned galaxies are in the main sequence, and only around $20$ and $15$ are, respectively, in the green valley and quiescent regimes. At intermediate mass, the quiescent regime already is the most populated by the galaxies with high kinematic misalignment, followed by the main sequence and the green valley, with percentages of approximately $45$, $30$, and $25$, respectively. The trend is completely reversed at high mass, where approximately $75$ per cent of misaligned galaxies are quiescent, and around $15.5$ and $5.5$ per cent are found, respectively, in the green valley and main sequence. Conversely, the kinematically aligned galaxies prefer the main sequence regime regardless of the stellar mass, with fractions that range from $85$ per cent at low mass, to $45$ at high mass. The decline in the latter fraction occurs at the expense of an increase in the fraction of galaxies in the green valley, and a much more modest increase in the fraction of galaxies in the quiescent regime. Note that we have verified that adopting alternative green valley definitions used in the literature \citep[e.g.,][]{Salim2014} leads to small variations in the fractions reported in Fig.~\ref{fig:fraction_vs_mass}, but does not alter the observed global trend. Likewise, the proximity of many high-mass misaligned galaxies to the lower boundary of the green valley implies larger uncertainties in the corresponding fractions. However, accounting for these effects does not change our main conclusion: massive misaligned galaxies tend to avoid regions of high star formation.

\subsection{Ionized gas}
\label{SS:ionized_gas}

In our sample of optical data, the most direct way to trace gas is through ionized hydrogen emission. To study this component, we first need to correct the \Ha\ emission maps for dust attenuation. The emission line maps provided in the MEGACUBES are corrected for Galactic extinction by default, but not for attenuation caused by the interstellar medium of the host galaxies. Therefore, we apply the following correction to the observed flux at wavelength $\lambda$, $F_\lambda$:
\begin{equation}
    F_{\lambda}^\prime = F_{\lambda}\, 10^{0.4\, A_{\lambda}}\;,
\label{eq:flux}    
\end{equation}
where $A_\lambda$ is the interstellar extinction at $\lambda$. This latter term is derived from the extinction in the V band, $A_V$, and the attenuation curve from \citet{Cardelli1989}, $k(\lambda)$, using
\begin{equation}
    A_{\lambda} = A_V\,k(\lambda)\;.
\label{eq:interstellar_extinction}
\end{equation}
We estimate $A_V$ from the observed Balmer decrement, adopting an intrinsic line ratio between \Ha\ and \Hb\ of 2.86 \citep[assuming case B recombination, electron density $n_e = 100\; {\rm cm^{-3}}$ and temperature $T_e = 10\,000$\; K;][]{Osterbrock2006}, the attenuation curve, and a ratio of total to selective extinction of $R_V = 3.1$:
\begin{equation}
    A_V = 7.22 \log\left(\dfrac{F_{\Ha}/F_{\Hb}}{2.86}\right)\;.
\label{eq:extinction}
\end{equation}
Attenuation corrected maps are derived using only those spaxels in which the \Ha\ and \Hb\ emission lines have a signal‑to‑noise ratio (S/N) higher than three. Spaxels that do not meet this requirement---where the Balmer decrement cannot be measured reliably---are masked and excluded from the analysis.

\begin{figure}
    \centering
	\includegraphics[width=\columnwidth]{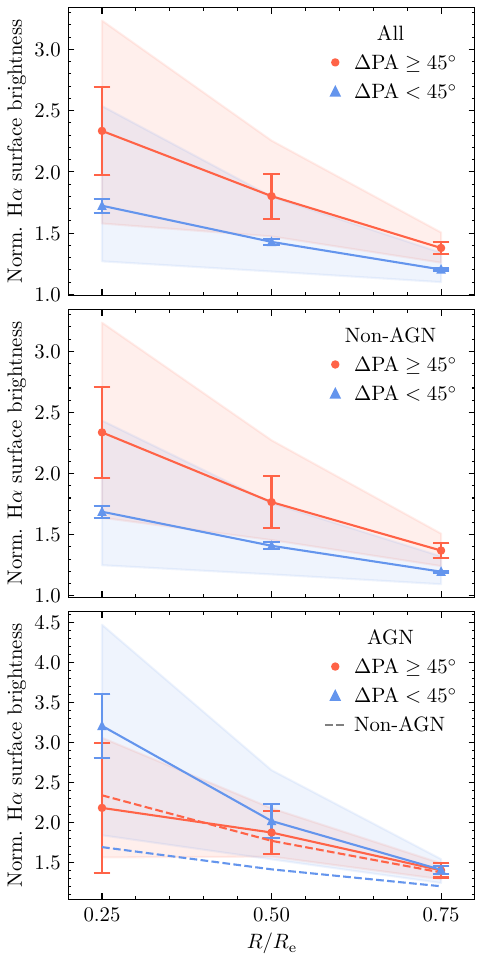}
    \caption{Enclosed \Ha\ luminosity as a function of radius, normalized to \Reff, of galaxies with high (red dots) and low (blue triangles) kinematic misalignment in the original sample (top panel), with no AGN (middle panel), and with AGN (bottom panel). Data points indicate the median, error bars show the $3\sigma$ uncertainty on the population median inferred from 1000 bootstrap resamples, and shaded regions show the interquartile range of the distributions. For reference, the dashed lines on the bottom panel correspond to the median values of the non-AGN subset shown on the middle panel.}
    \label{fig:concentration_profiles}
\end{figure}

Using these maps, we assess how concentrated the ionized gas is in kinematically misaligned galaxies compared to aligned ones. For each galaxy, we estimate the normalized H$\alpha$ surface brightness in the central region by taking the ratio of the $F_\Ha^\prime$ surface density within apertures of $0.25$, $0.5$, and $0.75$ \Reff\ to that within one \Reff. Note that we only determine the concentration for a galaxy if at least $75$ per cent of its projected area within each of the four apertures is covered by unmasked spaxels. The number of galaxies with high- and low-kinematic misalignment that satisfy this condition is, respectively, $172$ and $4016$. The top panel of Fig.~\ref{fig:concentration_profiles} shows the concentration of ionized gas for galaxies with high (red) and low (blue) kinematic misalignment across the probed apertures. This reveals that the ionized gas tends to be more centrally concentrated in misaligned galaxies. Despite some overlap between the two distributions, the observed differences are significant, as KS tests return $p$-values $\ll 0.001$ regardless of the aperture considered. As shown in Appendix~\ref{A:concentration}, this trend holds when controlling for stellar mass and light concentration, although at a lower statistical significance at intermediate and high masses.

To isolate the contribution of AGN to the central concentration of ionized gas, we further split the sample into non-AGN and AGN galaxies following \citet{Raimundo2023}, combining spatially resolved Baldwin, Phillips \& Terlevich (\citeyear{Baldwin1981}) diagrams with the multiwavelength classification by \citet{Comerford2024} to identify the dominant ionization source. This classification yields $142$ misaligned and $3798$ aligned galaxies in the non-AGN subset, and $30$ misaligned and $218$ aligned galaxies in the AGN subset\footnote{These figures correspond to an AGN fraction of $\approx 21$ per cent for misaligned galaxies and $\approx 6$ per cent for aligned galaxies. These fractions are fully consistent with previous findings which suggest that the physical conditions in misaligned galaxies are favourable to AGN \citep{Raimundo2023, Raimundo2025}.}. The middle panel of Fig.~\ref{fig:concentration_profiles} shows that the observed trend is not driven by AGN only, as higher concentrations of ionized gas are also observed in the non-AGN subset. Note that, since \Ha\ emission traces recent star formation \citep[$<20$ Myr;][]{Kennicutt1998}, this suggests that nuclear star formation is enhanced when galaxies are fuelled with misaligned gas. In contrast, in the AGN subset (bottom panel), misaligned galaxies exhibit concentrations more similar to those of their aligned counterparts, to the extent that we cannot rule out the possibility that both distributions are drawn from the same parent population, with KS test $p$-values $> 0.02$ across all three apertures. Interestingly, if we only considered kinematically aligned hosts, AGN show a statistically significant higher concentration compared to non-AGN (see the blue dashed line in the bottom panel of the figure), with KS test $p$-values $\ll 0.001$ across the three apertures. On the other hand, in kinematically misaligned galaxies, there is no significant difference between AGN and non AGN (compare the red dots with the red dashed line in the bottom panel), for which KS tests return large $p$-values ($\gtrsim 0.5$) for the three apertures. As discussed in Section~\ref{S:discussion}, we attribute this latter result to a mismatch between the duration of kinematic misalignments and AGN duty cycles. We note that the trends involving AGN do not seem to result from structural differences between the subsamples (see Appendix~\ref{A:concentration}).

Following \citet{Kim1989}, we now estimate the mass of ionized gas ($M_{\rm HII}$) in solar masses in each spaxel of our galaxies through:
\begin{equation}
    M_{\rm HII} = 2.8 \times 10^{17}\, D^{2}\, F'_{\Ha} \, n_e^{-1}\;,
\label{eq:mhii}
\end{equation}
where $D$ is the luminosity distance to the galaxies in Mpc (assuming our adopted cosmology), $F^\prime_\Ha$ is the extinction-free \Ha\ flux in units of ${\rm erg\;cm^{-2}\;s^{-1}}$ from eq.~(\ref{eq:flux}), and $n_e$ is the electron density in cm$^{-3}$ in each spaxel which we infer from the flux ratio between the lines $\Sii\lambda 6716$ and $\Sii\lambda 6731$ using equation (7) from \citet{Sanders2016}. We note that in several spaxels the line ratio lies above the saturation limit, corresponding to $n_e \lesssim 100\;{\rm cm^{-3}}$, where that equation becomes insensitive to the electron density. In these cases, we assign an upper limit of $100\;{\rm cm^{-3}}$. We mask those spaxels with a S/N lower than three in any of the emission lines needed for the calculation of the ionized gas mass (i.e., \Ha, \Hb, $\Sii\lambda 6716$ and $\Sii\lambda 6731$).

\begin{figure}
    \centering
	\includegraphics[width=\columnwidth]{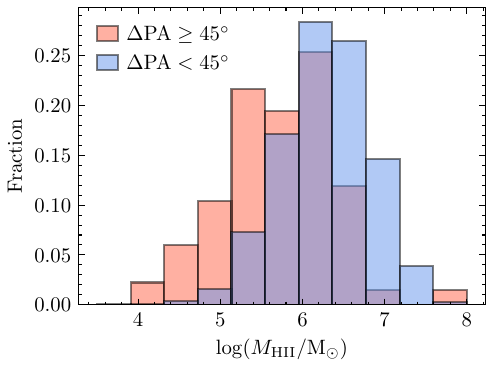}
    \caption{Distribution of total ionized gas mass within the \Reff\ of galaxies with high (red) and low (blue) kinematic misalignment. The histograms show, for each subset, the fraction of galaxies in a given bin so each distribution is normalized to unity. The observed differences are statistically significant according to a two-sample KS test, with $p{\rm -value} \ll 0.001$.}
    \label{fig:ionized_gas_mass_hist}
\end{figure}

Finally, we estimate the total ionized gas mass for each galaxy by integrating its value across all the spaxels within \Reff. The histograms in Fig.~\ref{fig:ionized_gas_mass_hist} show the distributions of $ M_{\rm HII}$ for kinematically misaligned (red) and aligned (blue) galaxies. Again, in this figure, we only include galaxies with at least $75\%$ of their projected effective area covered by unmasked spaxels. The reservoir of ionized gas is on average less massive in misaligned galaxies, in line with the stellar mass and SSFR distributions (Fig.~\ref{fig:ssfr_mass_hist}), with a median value of $10^{5.8}\; {\rm M_ \odot}$ compared to $10^{6.3}\; {\rm M_ \odot}$ for their aligned counterparts. A two-sample KS test applied on the distributions returns a $p$-value $\ll 0.001$, supporting the observed difference. This difference also holds if we split the sample into the three bins of stellar mass. The range of values we find ($\sim 10^{4}-10^{8}\;M_\odot$) is consistent with those reported by \citet{Nascimento2019} for a high-mass sample of AGN and non-AGN controls from MaNGA.

\subsection{Radial profiles}
\label{SS:radial_profiles}

In this section, we assess the role of kinematically misaligned accretion on the local star formation activity as a function of stellar mass. To do so, we divide the sample into the three stellar mass ranges defined in Section~\ref{SS:mstar_ssfr}, and build radial profiles of SSFR and ionized gas mass surface density ($\Sigma_{\rm HII}$). These profiles are obtained by grouping the spaxels, of all the galaxies, into six circularized radial bins (annuli), from $0$ to $1.5$ \Reff\ using a step of $0.25$ \Reff, and determining the median values of each property within each bin, as in \citet{Bluck2020b}.

\subsubsection{Specific star formation rate profiles}
\label{SSS:ssfr_profiles}

The SSFR profiles are derived from spatially resolved SSFR maps, which we obtain for each galaxy by dividing the SFR (averaged over the last $20$ Myr) in each spaxel by its corresponding stellar mass, provided that neither quantity is masked in the MEGACUBES data.

\begin{figure*}
    \centering
	\includegraphics[width=\textwidth]{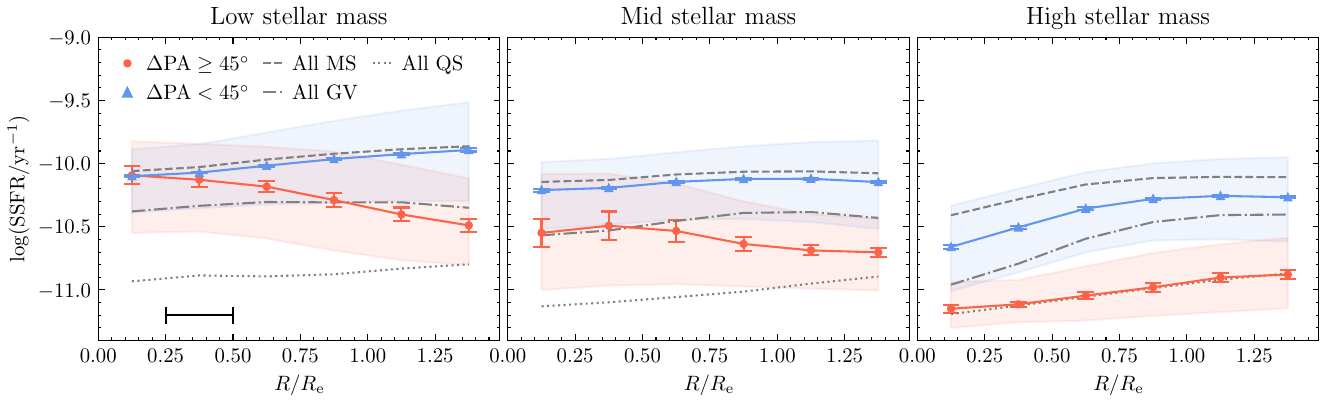}
    \caption{Median SSFR radial profiles of galaxies with high (red dots) and low (blue triangles) kinematic misalignment in the same three bins of stellar mass as in Fig.~\ref{fig:fraction_vs_mass}. The error bars show the $\mathbf{1 \sigma}$ uncertainty on the population median inferred from 1000 bootstrap resamples, and the shaded regions show the interquartile range of the profile distributions. The dashed, dot-dashed, and dotted profiles correspond, respectively, to all the MaNGA galaxies in the main sequence (MS), green valley (GV), and quiescent (QS) regimes. The horizontal black line in the bottom left corner of the first panel indicates the size of the radial bins.}
    \label{fig:ssfr_profiles}
\end{figure*}

In Fig.~\ref{fig:ssfr_profiles}, we show the SSFR radial profiles for galaxies with high and low kinematic misalignment (represented, respectively by red dots and blue triangles), grouped into the three stellar mass bins. Overall, the SSFR values of galaxies with high kinematic misalignment are lower than those of their low misalignment counterparts---particularly at high stellar mass---in agreement with the global trends observed in Fig.~\ref{fig:ssfr_mass_hist}. Only at low stellar masses do the central regions of highly misaligned galaxies show SSFR values comparable to those of galaxies with low misalignment. Despite these differences, the median SSFR in both subsets declines with increasing stellar mass at all radii. These results suggest that the global decline of the SSFR with the stellar mass observed in Fig.~\ref{fig:ssfr_mass} is mirrored at local scales in both subsets. However, the radial behaviour differs significantly when the gas is kinematically misaligned. At low stellar mass, the SSFR profile of highly misaligned galaxies decreases inside out. At intermediate mass, the profile flattens, and at high mass, it shows an inside-out increase. In contrast, galaxies with low kinematic misalignment display inside-out increasing SSFR profiles across all mass bins, with the steepest gradient observed in the highest mass bin. As discussed in Appendix~\ref{A:concentration}, the observed differences between the profiles of aligned and misaligned systems are not uniquely driven by differences in their underlying stellar structure.

The SSFR profiles of all MaNGA galaxies in the main sequence\footnote{To ensure we only use regular star-forming galaxies for this comparison, we restrict the main sequence subset to galaxies with $\log({\rm SSFR_{MS}}) = [-0.5, 0.5)$ as defined in Section~\ref{SS:mstar_ssfr}.}, green valley, and quiescent regimes shown in Fig.~\ref{fig:ssfr_profiles} provide additional reference points for comparing the effect of the misaligned accretion onto the local star formation activity. The similarity between the SSFR profiles of galaxies with low kinematic misalignment and those of the main sequence subset can be attributed to the predominance of main sequence systems in this group (see Fig.~\ref{fig:fraction_vs_mass}). However, at low and intermediate stellar masses, the SSFR profile of kinematically misaligned galaxies lies between those of the main sequence and quiescent subsets, exhibiting a distinct shape---even differing from the green valley profile---despite the dominant contribution of main sequence and quiescent systems in these mass ranges. In contrast, at high stellar masses, the profile closely resembles that of quiescent galaxies. 

\subsubsection{Ionized gas mass surface density profiles}
\label{SSS:mhii_profiles}

Using the MaNGA pixel scale of $0.5\,{\rm arcsec/pixel}$, we convert the ionized gas mass within each spaxel, obtained from eq.~\eqref{eq:mhii}, to a surface density in units of solar mass per kpc$^2$ as:
\begin{equation}
\Sigma_{\rm HII} = C(i)\,\frac{M_{\rm HII}}{(0.5\,D_A)^2}\;,
\label{eq:surface_density}    
\end{equation}
were $D_A$ is the angular diameter distance in kpc, so the denominator represents the physical area covered by one spaxel. The factor $C(i) = \cos(i) \approx b/a$ is introduced to correct the surface density for each galaxy's inclination, $i$, estimated from the axis ratio $b/a$ provided in the NSA catalogue.

\begin{figure*}
    \centering
	\includegraphics[width=\textwidth]{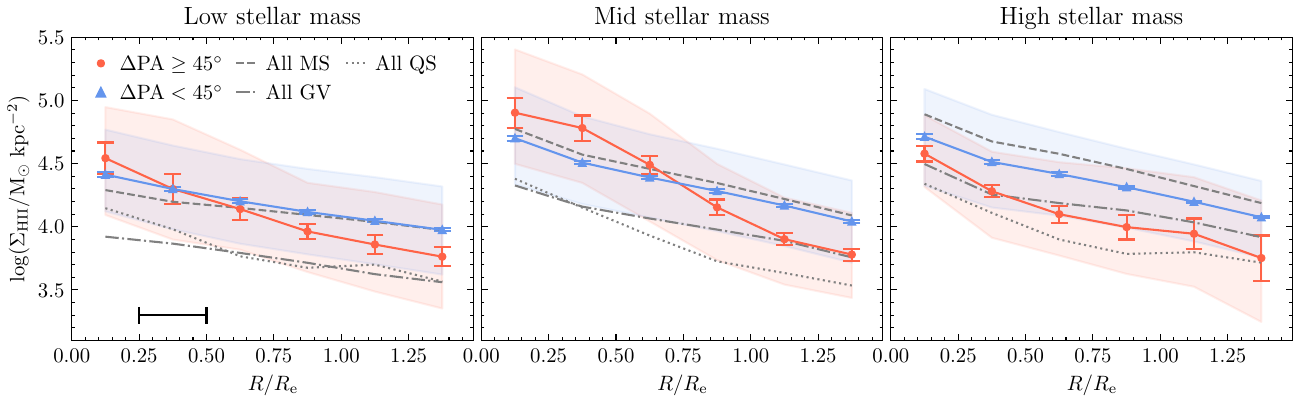}
    \caption{Same as Fig.~\ref{fig:ssfr_profiles} but for the ionized gas mass surface density profiles.}
    \label{fig:mhii_sd_profiles}
\end{figure*}

Fig.~\ref{fig:mhii_sd_profiles} shows the ionized gas mass surface density profiles for galaxies with low and high kinematic misalignment, grouped in the three bins of stellar mass. All profiles decline with radius regardless of stellar mass but galaxies with high kinematic misalignment, which exhibit more centrally concentrated ionized gas (Fig.~\ref{fig:concentration_profiles}), are characterised by steeper negative gradients of approximately one dex across the probed radius. At low and intermediate stellar masses, these steeper gradients result in higher central surface densities but lower values in the outermost radial bins compared to their more aligned counterparts or the main sequence subset. In the high-mass bin, however, the entire profile of misaligned galaxies lies below that of the aligned ones, exhibiting similar central values to green valley galaxies but lower values in the outskirts, comparable to those observed for the quiescent subset. Interestingly, the steepest profiles are found in intermediate-mass misaligned galaxies, driven by elevated central surface densities, with typical values around $10^{4.9}\;{\rm M_\odot \; kpc^{-2}}$. However, the profiles of misaligned galaxies are more similar at both low and high stellar masses. In contrast, galaxies with low kinematic misalignment display shallower gradients of approximately $0.5$ dex, with profile normalization increasing from the low- to intermediate-mass bin and remaining roughly constant between the intermediate- and high-mass bins. Note that, as shown in Appendix~\ref{A:SFR_profiles}, the use of upper limits for some $n_e$ is justified by the fact that we obtain consistent results from SFR surface density ($\Sigma_{\rm SFR}$) profiles---including a significant peak in $\Sigma_{\rm SFR}$ also at intermediate stellar mass.

The behaviour of the $\Sigma_{\rm HII}$ profiles of the galaxies with high kinematic misalignment contrasts with that of the SSFR profiles (Fig.~\ref{fig:ssfr_profiles}). Despite being characterised by lower SSFR, our results indicate that, in their central regions, kinematically misaligned galaxies exhibit comparable, if not higher, ionized gas mass surface densities to the aligned counterparts. Since the ionized gas traces ongoing star formation (see the corresponding $\Sigma_{\rm SFR}$ profiles in Fig.~\ref{fig:sfr_sd_profiles}), this result suggests that the observed decline in the SSFR of kinematically misaligned galaxies must be more related with an increase in the stellar mass surface density rather than a depressed $\Sigma_{\rm SFR}$ (see Section~\ref{S:discussion}).

\section{Discussion}
\label{S:discussion}

Longer relaxation timescales of misaligned gas structures are often associated with smaller native gas reservoirs in galaxies, which are less efficient at dissipating angular momentum through shocks \citep{Khim2021, Ristea2022, Baker2025}. This is one of the key arguments used to explain the higher prevalence of such structures in ETGs compared to late-type galaxies \citep{Davis2011, Chen2016, Bryant2019, Ristea2022, Raimundo2023}. This applies to our sample, where most misaligned galaxies exhibit a Sérsic index greater than two, as well as larger central stellar velocity dispersion than aligned galaxies of similar mass (see Appendix~\ref{A:morphology}). Both characteristics suggest the presence of a bulge component typical of early‑type morphologies \citep[e.g.,][]{HDS2022}, and possibly reflect a history of substantial quenching. Therefore, the central enhancement of ionized gas we observe in most misaligned galaxies is consistent with an increase in nuclear star formation triggered by the recent accretion of fresh gas.

Our results are in qualitative agreement with previous works that find that the fraction of galaxies with kinematically misaligned gas structures sharply declines above $\approx 10^{10.5}\;{\rm M_\odot}$ \citep{Davis2011, Jin2016, Zhou2022}, and that these structures promote central star formation \citep{Chen2016, Xu2022}. Yet, our analysis provides significant evidence that the accretion of misaligned gas has a different impact on the star formation activity of galaxies depending on their stellar mass. At low mass, these events are associated with main-sequence levels of star formation activity in the nucleus. At higher masses, most kinematic misalignments are observed in green valley and quiescent galaxies. Nevertheless, the ionized gas mass and SFR surface density profiles show that these massive galaxies exhibit ongoing central star formation ($< 20$ Myr), at higher rates than typical for these regimes. Therefore, in massive galaxies, these events likely indicate rejuvenation via external accretion, although with a small impact on the SSFR of the host.

\begin{figure*}
    \centering
	\includegraphics[width=\textwidth]{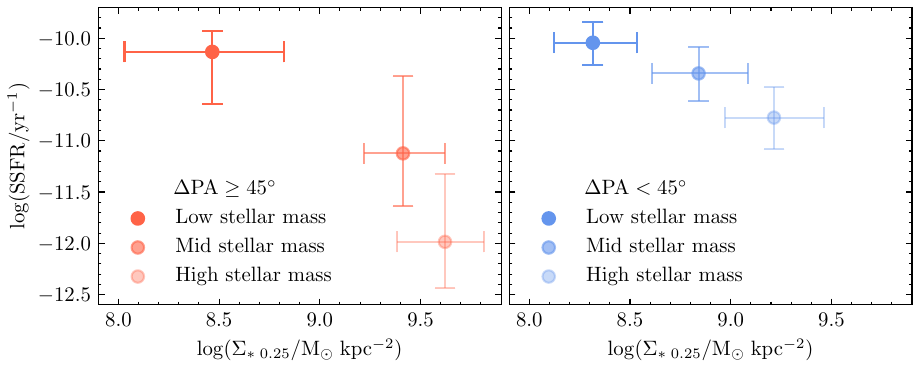}
    \caption{Global specific star formation rate vs stellar mass surface density within the central $0.25$ \Reff\ as a function of global stellar mass of galaxies with high (left) and low (right) kinematic misalignment. In both panels, the different tones of the markers indicate, from darker to lighter, increasing global stellar mass corresponding to the same mass bins of Fig.~\ref{fig:fraction_vs_mass}. The coordinates of the data points are determined from the median of the SSFR distribution of the galaxies in each mass bin and the median of the stellar mass surface density distribution of all their corresponding central spaxels. The error bars show the interquartile ranges of the two distributions. Kinematically misaligned massive galaxies are compact and therefore the contribution of newly formed stars is modest compared to pre-existing mass, resulting into low SSFRs.}
    \label{fig:ssfr_vs_compactness}
\end{figure*}

The spatial distribution of stellar mass in intermediate- and high-mass misaligned galaxies explains their low SSFRs despite ongoing central star formation. Fig.~\ref{fig:ssfr_vs_compactness} shows the median values of global SSFR and stellar mass surface density within the central $0.25\;$\Reff\ ($\Sigma_{*\; 0.25}$) as a function of global stellar mass for galaxies with low and high kinematic misalignment. Note that we use $\Sigma_{*\; 0.25}$ as a proxy for compactness \citep{Zolotov2015}. When plotted against each other in a log-log space, the SSFR as a function of $\Sigma_{*\; 0.25}$ of misaligned galaxies describes a characteristic knee where the former parameter plummets more than two orders of magnitude\footnote{The drop in SSFR could be even stronger, as these values represent upper limits in the extremely low star formation regime \citep{Salim2016}.} as compactness increases. This knee is persistently observed across a wide range of redshifts in samples of real galaxies \citep{Cheung2012, Fang2013, Barro2017, Huertas-Company2018, Wang2018}, resembling the final compaction events driving the "blue nugget" phase that precedes inside-out quenching in high redshift galaxies predicted by cosmological simulations \citep{Zolotov2015, Tachella2016, Lapiner2023}. In contrast, for the aligned subset of galaxies, $\log({\rm SSFR})$ and $\log(\Sigma_{*\; 0.25})$ span much narrower ranges of higher SSFR and lower compactness, following an approximately linear anticorrelation of slope $\lesssim -1$. Because most misaligned massive galaxies are compact, centrally concentrated star formation contributes only modestly to the stellar mass growth relative to the pre-existing stellar content, resulting in the observed low SSFRs. This is consistent with the idea that rejuvenation events contribute a small fraction to the total stellar mass of the host \citep{Chauke2019, Tanaka2024}.

As discussed in Appendix~\ref{A:concentration}, a more concentrated mass profile alone cannot fully account for the enhanced central gas concentration observed in galaxies with strong kinematic misalignment. However, the relative importance of kinematic misalignment compared to the underlying structure of the host galaxy appears to depend on stellar mass. At low masses, where galaxies are less compact, the shocks and stellar torques associated with kinematic misalignment may play a key role in funnelling gas towards the nucleus. As stellar mass increases, the gravitational potential becomes the dominant factor, largely determining how newly accreted gas is distributed across the galaxy, with kinematic misalignments likely playing a secondary role. 

According to \citet{Baker2025}, in the EAGLE simulation \citep{Schaye2015} most unstable kinematic misalignments are short in duration, meaning that configurations with ${\rm \Delta PA} \approx 0^\circ$ and ${\rm \Delta PA} \approx 180^\circ$ predominate in galaxies at $z=0$. This is consistent with our sample, where $58$ per cent of misaligned galaxies show ${\rm \Delta PA} > 135^\circ$. In addition, these authors find that long-duration unstable misalignments preferentially occur in gas-poor massive galaxies, and suggest smooth gas accretion (e.g. supplied by halo cooling) as a possible explanation. This is also consistent with our sample, where the high-mass bin accumulates the highest fraction of unstable configurations (i.e., $45^\circ \leq$ \DPA~$< 135^\circ$) which represent more than $60$ per cent of the misalignments in that bin, suggesting longer relaxation times in more massive systems. However, taking compactness as a proxy for bulge growth \citep{Fang2013}, this trend might also be driven by the tendency of our massive misaligned galaxies to exhibit more compact stellar mass distributions compared to low-mass galaxies (see Figs.~\ref{fig:ssfr_vs_compactness} and~\ref{fig:c95}, and Appendix~\ref{A:morphology}). The gravitational potential associated with such distribution exerts weaker torques on misaligned gas, allowing it to remain in an unstable configuration for longer \citep{Tholine1982, Lake1983}.

Most misaligned galaxies in our sample show high central concentrations of ionized gas, indicating that a substantial reservoir is present even when nuclear activity is not detected (see bottom panel of fig.~\ref{fig:concentration_profiles}). Our radial profiles of ionized gas mass, and SFR surface density, show---for the first time---that the central accumulation of gas driven by misaligned structures is enhanced in galaxies with stellar masses between $10^{10} - 10^{10.6}\;{\rm M_\odot}$. This mass range roughly corresponds to a halo mass of $\sim 10^{12}\;{\rm M_\odot}$ above which supernova feedback stops being effective at quenching galaxies \citep{Shankar2006, Dekel2006, Dekel2019}. Since misaligned gas tends to accumulate in the central region of galaxies and supernova ejecta are locked in massive halos, such galaxies could start building a persistent gas reservoir capable of fuelling their SMBHs over a sustained period. However, AGN duty cycles are short ($\sim 10^5\; {\rm yr}$; \citealt{Schawinski2015}) compared to the typical relaxation time of misaligned gas ($> 10^8\;{\rm yr}$; \citealt{Baker2025}). Therefore, not all misaligned galaxies are expected to be active at the current epoch despite having fuel available. Moreover, even when nuclear activity occurs, its luminosity may remain low if, for example, the weaker torques associated with the mass distribution of massive misaligned galaxies translate into reduced gas inflow rates and, ultimately, lower black hole accretion rates. Thus, low inflow rates in a mass range where SMBH growth is expected \citep{Dekel2019} could explain the lack of correlation between AGN luminosity and stellar mass reported by \citet{Winiarska2025}.

\section{Summary and conclusions}
\label{S:conclusion}

In this paper, we have studied the global and local star formation activity, as well as the ionized gas content, of kinematically misaligned galaxies across stellar mass. Using a sample of $201$ galaxies exhibiting kinematic misalignment between gas and stars, and a comparison sample of $3689$ aligned galaxies, both from MaNGA, we have characterised their global specific star formation rate, central concentration and total mass of ionized gas, as well as the radial distribution of these parameters. The main findings of this study are outlined below.

\begin{itemize}

    \item Kinematically misaligned galaxies tend to be less massive and show lower specific star formation rates than aligned galaxies, supporting the idea that misaligned gas structures persist longer in systems with smaller native gas reservoirs and reduced angular momentum dissipation via shocks (Fig.~\ref{fig:ssfr_mass_hist}).
    
    \item Kinematically misaligned galaxies avoid the region in the parameter space of high specific star formation rate and stellar mass, which is densely populated by aligned galaxies and the broader SDSS sample (Fig.~\ref{fig:ssfr_mass}). This absence of massive misaligned galaxies in the star-forming main sequence suggests that misaligned accretion can only sustain main-sequence levels of star formation in low-mass systems.
   
    \item Ionized gas is more centrally concentrated in kinematically misaligned galaxies than in aligned systems (Fig.~\ref{fig:concentration_profiles}). This enhancement is primarily driven by kinematic misalignments at low stellar masses, while at intermediate and high masses the underlying mass distribution dominates, with misalignment providing a potential enhancement factor (Fig.~\ref{fig:concentration_profiles_c95}).
    
    \item In misaligned galaxies, AGN and non-AGN exhibit similar concentrations of ionized gas (Fig.~\ref{fig:concentration_profiles}). This suggests that misaligned gas is replenishing the nuclear reservoir and that all misaligned galaxies may have the necessary fuel to power their AGN. The absence of AGN in some misaligned galaxies is likely due to AGN flickering occurring on timescales up to three orders of magnitude shorter than the lifetime of the misalignment.
    
    \item The typical central density of star-forming gas in misaligned galaxies peaks at $M_* = 10^{10}-10^{10.6}\;{\rm M_\odot}$, with values higher than those observed in aligned or main-sequence galaxies (Fig.~\ref{fig:mhii_sd_profiles}). Curiously, this stellar mass range coincides with the halo mass above which supernova feedback stops being effective at quenching galaxies (last paragraph of Section~\ref{S:discussion}).
    
    \item Above a stellar mass of $\approx 10^{10}\; {M_\odot}$, most kinematically misaligned galaxies are compact, with central stellar surface densities $> 10^{9.2}\;{\rm M_\odot\; kpc^{-2}}$. In contrast, the majority of aligned galaxies are less compact ($\Sigma_{*\; 0.25} < 10^{9.5}\;{\rm M_\odot\; kpc^{-2}}$), regardless of stellar mass (Fig.~\ref{fig:ssfr_vs_compactness}). The high compactness of the misaligned massive galaxies likely explains their low SSFRs despite enhanced central star formation. 

    \item In agreement with cosmological simulations, most kinematic misalignments in our sample likely originate from short-lived unstable configurations. Unstable kinematic misalignments are preferentially found in massive galaxies, where they represent more than $60$ per cent of the misalignments (second to last paragraph of Section~\ref{S:discussion}). This suggests that unstable misaligned structures persist longer in massive galaxies, likely because the centrally concentrated mass distribution of these systems contributes to extending the relaxation timescale.
    
\end{itemize}

Altogether, our results support the established picture where misaligned accretion drives gas toward the centre of galaxies, refilling the gas reservoir that feeds central mass growth and nuclear activity \citep{Raimundo2013, Chen2016, Xu2022, Raimundo2023, Raimundo2025}. On a more global galactic scale, these accretion events seem to be an important source of star formation---particularly in ETGs---highlighting the role of external interactions in the recent star formation history of quiescent systems.

\section*{Acknowledgements}
The authors thank the referee for their insightful suggestions, which have helped to improve the presentation of our results. This work was supported by the Science and Technology Facilities Council (STFC) of the UK Research and Innovation via grant reference ST/Y002644/1. RR acknowledges support from  Conselho Nacional de Desenvolvimento Cient\'{i}fico e Tecnol\'ogico  (CNPq, Proj. CNPq-445231/2024-6,311223/2020-6, 404238/2021-1, and 310413/2025-7), Funda\c{c}\~ao de amparo \`{a} pesquisa do Rio Grande do Sul (FAPERGS, Proj. 19/1750-2 and 24/2551-0001282-6) and Coordena\c{c}\~ao de Aperfei\c{c}oamento de Pessoal de N\'{i}vel Superior (CAPES, 88881.109987/2025-01). M.V. gratefully acknowledges financial support from the Independent Research Fund Denmark (grant 3103-00146) and from the Carlsberg Foundation (grant CF23-0417). J.L. would like to extend his thanks to Shenli Tang for helpful discussions about AGN ionization, and to Matt Malkan for reading the first draft and providing encouraging feedback. This research has made use of MaNGA data from SDSS IV. Funding for the Sloan Digital Sky Survey IV has been provided by the Alfred P. Sloan Foundation, the U.S. Department of Energy Office of Science, and the Participating Institutions. SDSS acknowledges support and resources from the Center for High-Performance Computing at the University of Utah. The SDSS web site is \url{www.sdss4.org}.

\section*{Data availability}
The MaNGA MEGACUBES used in this work are publicly available through a web interface at \url{https://manga.linea.org.br/} and \url{https://manga.if.ufrgs.br/}. The catalogue of global properties used in this work is available at \url{https://salims.pages.iu.edu/gswlc/}, and the additional maps from the MaNGA data-reduction pipeline at \url{https://data.sdss.org/sas/}. The data underlying this article will be shared on reasonable request to the corresponding author.


\bibliographystyle{mnras}
\bibliography{biblio}



\appendix

\section{Disentangling the effect of kinematic misalignments from structural concentration}
\label{A:concentration}

The stellar mass distribution of a galaxy sets the local gravitational potential, and during accretion events, newly accreted gas is expected to respond to this potential and settle into a similar radial configuration. Misaligned galaxies typically exhibit early-type morphologies (Appendix~\ref{A:morphology}), characterised by higher structural concentrations relative to aligned systems. This motivates the need to disentangle whether the trends observed in the distribution of gas and star formation arise from kinematic misalignment, or instead reflect the underlying structural properties of the host galaxies.

\begin{figure}
    \centering
	\includegraphics[width=\columnwidth]{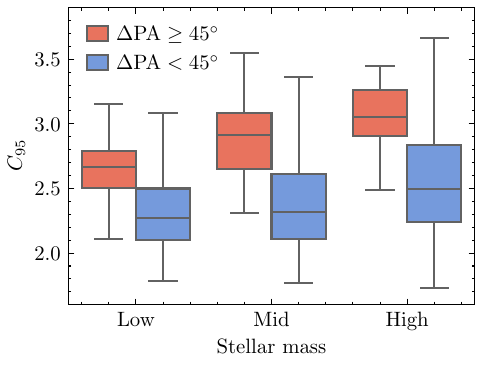}
    \caption{Box plots summarizing the distribution of light concentration for galaxies with high (red) and low (blue) kinematic misalignment within the same stellar mass bins as in Fig.\ref{fig:fraction_vs_mass}. The boundaries of the boxes indicate the interquartile ranges of the data, and central lines show the median values.}
    \label{fig:c95}
\end{figure}

In this appendix, we use the concentration of light as a proxy for structural concentration. We adopt the ratio of the radii enclosing $90$ and $50$ per cent of the $r$-band light, $C_{95} = R_{90}/R_{50}$, with both quantities taken from the NSA catalogue. The red boxes in Fig.~\ref{fig:c95} show the $C_{95}$ distribution for galaxies with high kinematic misalignment across the three stellar mass bins defined in Section~\ref{SS:mstar_ssfr}, while the blue boxes show the corresponding distributions for aligned systems. At fixed stellar mass, misaligned galaxies are more concentrated systems than their aligned counterparts.

To isolate the effect of kinematic misalignment on the distribution of gas and star formation, we construct, within each stellar mass bin, samples of misaligned and aligned galaxies matched in $C_{95}$. For each misaligned galaxy in a given subset, we select candidate aligned controls with $\mid\Delta C_{95}\mid\; < 0.1$. Aligned galaxies without matches within this threshold are excluded from the analysis. Next, all potential pairs are ranked by their absolute separation in $C_{95}$ and assigned using a global greedy algorithm that prioritises the closest matches. When multiple misaligned galaxies compete for the same control, it is assigned to the pair with the smallest $\mid\Delta C_{95}\mid$, so each misaligned galaxy can be matched with up to five unique controls. To assess the quality of the match in a given subset of galaxies, we keep track of the match ratio ($r_m$), defined as the average number of controls per misaligned galaxy, and the number of discarded misaligned systems with no matches ($n_0$).

\subsection{Concentration of gas}
\label{AA:gas_concentration_c95}

Here we examine the gas concentration as a function of stellar mass for the same subsets of galaxies presented in Fig.~\ref{fig:concentration_profiles}, but restricting the comparison to systems matched in $C_{95}$. For details on the estimation of gas concentration, see section~\ref{SS:ionized_gas}.

\begin{figure*}
    \centering
	\includegraphics[width=\textwidth]{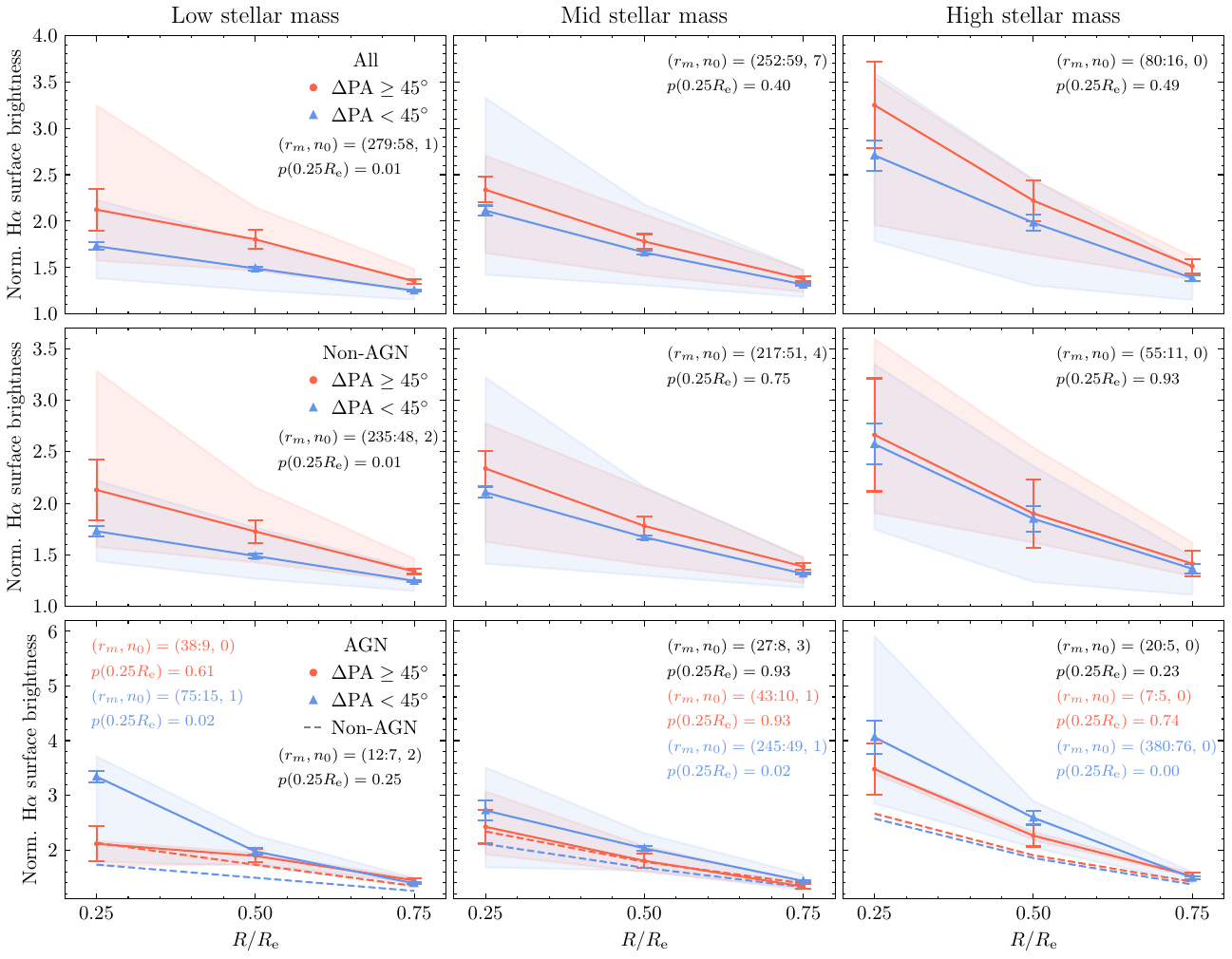}
    \caption{Enclosed \Ha\ luminosity as a function of radius, normalized to \Reff, for subsamples of kinematically misaligned galaxies (red dots) and aligned control systems (blue triangles) matched in $C_{95}$. From top to bottom, rows correspond to the full sample, the non-AGN, and the AGN subsamples. Within each column, galaxies are grouped into the same stellar mass bins as in Fig.~\ref{fig:fraction_vs_mass}. Data points show the median, with error bars indicating the $1\sigma$ uncertainty on the median estimated from $1000$ bootstrap resamples including re-matching, and shaded regions representing the interquartile range. Dashed lines in the bottom panels reproduce the median profiles of the non-AGN subsample shown in the corresponding middle panels. Each panel reports the match ratio and the number of galaxies without matches that have been excluded from this analysis ($r_m$, $n_0$), along with the $p$-value of a two-sample KS test comparing the distributions within the smallest aperture. The bottom row additionally presents these quantities for the comparison between AGN and non-AGN subsamples in both misaligned (red) and aligned (blue) systems.}
    \label{fig:concentration_profiles_c95}
\end{figure*}

The outcome of this matched analysis is shown in Fig.~\ref{fig:concentration_profiles_c95}, which presents the enclosed \Ha\ luminosity within $0.25$, $0.5$, and $0.75$ \Reff, normalised by the value within one \Reff, for galaxies with high (red dots) and low (blue triangles) kinematic misalignment. The top row shows that, overall, misaligned galaxies tend to exhibit more centrally concentrated ionised gas than structurally matched aligned systems across all stellar masses. This difference is statistically significant at low mass ($p = 0.01$), but becomes less significant towards higher masses, although at intermediate mass the median values still differ by $\gtrsim 1\sigma$. A similar trend is observed when excluding AGN hosts (middle row). In contrast, the bottom row shows a reversal for AGN hosts, with aligned galaxies displaying, on average, higher gas concentrations. However, this result is of low statistical significance, as indicated by the large $p$-values displayed in the figure and the overlap of the error bars. Finally, the comparison between AGN and non-AGN subsamples (solid versus dashed lines in the bottom panels) suggests that, in aligned galaxies, higher central gas concentrations are required to power SMBH activity, particularly at high stellar masses. Conversely, in misaligned systems, both active and inactive galaxies appear to host similarly concentrated central gas reservoirs. We caution, however, that the subset of misaligned AGN is small, and these results should therefore be interpreted with care.

From these results, we conclude that kinematic misalignments could be particularly important in driving the enhanced central gas concentration observed in low-mass galaxies ($M_* \leq 10^{10} {\rm M_\odot}$). In contrast, in more massive systems, the increased central gas concentration appears to be primarily governed by the shape of the host galaxy’s gravitational potential, with kinematic misalignments playing a secondary role.

\subsection{Specific star formation rate profiles}
\label{AA:ssfr_profiles_c95}

The higher structural concentration of kinematically misaligned galaxies implies that their central regions are intrinsically denser compared to aligned galaxies (Fig.~\ref{fig:ssfr_vs_compactness}). The existence of a spaxel-level correlation between local stellar mass surface density and star formation rate \citep{CanoDiaz2016, Hsieh2017} might produce some differences in the SSFR profiles unrelated to the kinematic misalignments. Here, we utilise the samples matched in $C_{95}$ to identify the contribution of kinematic misalignments to the SSFR profiles.

\begin{figure*}
    \centering
	\includegraphics[width=\textwidth]{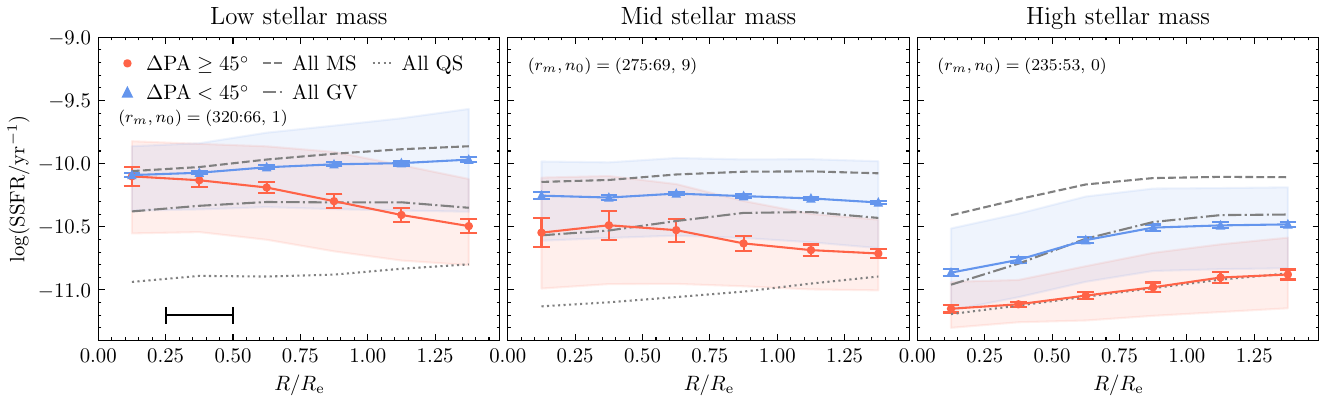}
    \caption{Median SSFR radial profiles of galaxies with high (red dots) and low (blue triangles) kinematic misalignment and aligned control systems (blue triangles) matched in $C_{95}$ in the three mass bins defined in Fig.~\ref{fig:fraction_vs_mass}. Data points show the median, with error bars indicating the $1\sigma$ uncertainty on the median estimated from $1000$ bootstrap resamples including re-matching, and shaded regions representing the interquartile range.The dashed, dot-dashed, and dotted profiles correspond, respectively, to all the MaNGA galaxies in the main sequence (MS), green valley (GV), and quiescent (QS) regimes. Each panel reports the match ratio and the number of galaxies without matches that have been excluded from this analysis ($r_m$, $n_0$). The horizontal black line in the bottom left corner of the first panel indicates the size of the radial bins.}
    \label{fig:ssfr_profiles_c95}
\end{figure*}

Fig.~\ref{fig:ssfr_profiles_c95} shows the SSFR profiles of misaligned galaxies (red dots) and their aligned control systems (blue triangles) in the three mass bins defined in Section~\ref{SS:mstar_ssfr}. Reference profiles for all MaNGA galaxies in the main sequence, green valley, and quiescent regimes are represented by dashed, dot-dashed, and dotted lines. While the profiles of misaligned galaxies are the same as the ones displayed in Fig.~\ref{fig:ssfr_profiles} (because the misaligned sample is the same), the profiles of the aligned controls look slightly flatter. At low and intermediate stellar mass, the relative change in the SSFR values of the central region of aligned control galaxies is smaller compared to the outer parts, whereas at high mass, the profile preserves the inside-out increasing shape but with lower values. Nevertheless, the main differences observed for the full sample (Fig.~\ref{fig:ssfr_profiles}) persist across stellar mass after $C_{95}$ is held fixed: galaxies with high kinematic misalignment have generally lower SSFR profiles than aligned ones, although their central regions are similar at low mass.

\section{Star formation rate surface density profiles}
\label{A:SFR_profiles}
This appendix shows the SFR surface density profiles of the subsets investigated in this work. These profiles have been derived from SFR maps from the MEGACUBES following the same procedure described in Section~\ref{SS:radial_profiles}, replacing the numerator of eq.~(\ref{eq:surface_density}) by the SFR of each spaxel.

\begin{figure*}
    \centering
	\includegraphics[width=\textwidth]{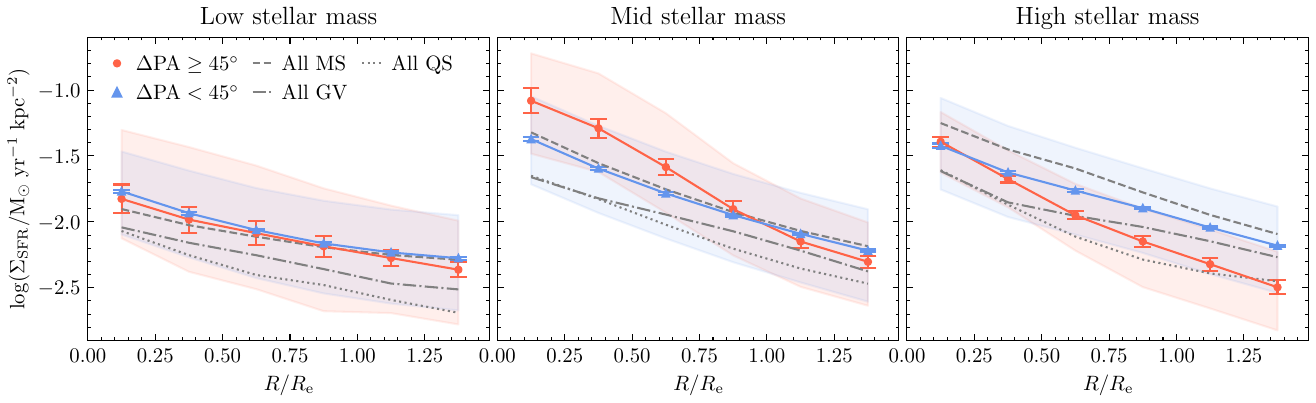}
    \caption{Same as Fig.~\ref{fig:ssfr_profiles} but for the SFR surface density derived from stellar population synthesis by \citet{Riffel2023}.}
    \label{fig:sfr_sd_profiles}
\end{figure*}

The $\Sigma_{\rm SFR}$ profiles are shown in Fig.~\ref{fig:sfr_sd_profiles}. Red circles and blue triangles indicate the median SFR surface density as a function of galactocentric distance for galaxies with, respectively, high and low kinematic misalignment across the three stellar mass bins defined in Section~\ref{SS:mstar_ssfr}. Reference profiles for all MaNGA galaxies in the main sequence, green valley, and quiescent regimes are represented by dashed, dot-dashed, and dotted lines. These profiles reproduce the trends seen in the $\Sigma_{\rm HII}$ profiles (Fig.~\ref{fig:mhii_sd_profiles}): misaligned galaxies are characterised by stronger negative radial gradients than aligned counterparts, particularly at intermediate and high stellar masses. Their central SFR surface densities are comparable to, or even exceed, those of aligned galaxies, with the highest values occurring at intermediate mass. Finding consistent trends by using SFRs derived from stellar population synthesis and ionized gas masses from \Ha\ flux highlights the robustness of our results.

\section{Morphological characterisation}
\label{A:morphology}
Here, we assess the morphology of our sample. To do so, we use measurements of the Sérsic index and the axis ratio, $b/a$, from the NASA-Sloan Atlas catalogue, and we estimate the central stellar velocity dispersion from the MEGACUBES.

\begin{figure}
    \centering
	\includegraphics[width=\columnwidth]{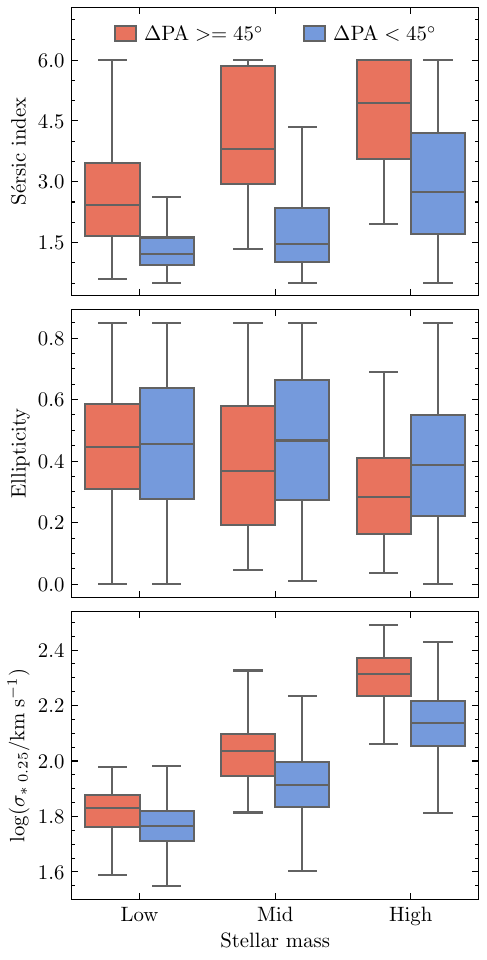}
    \caption{Box plots summarizing the distribution of morphological parameters for galaxies with high (red) and low (blue) kinematic misalignment within the same stellar mass bins as in Fig.\ref{fig:fraction_vs_mass}. From top to bottom, the panels display the distributions of Sérsic index, observed ellipticity, and central stellar velocity dispersion measured witin $0.25$ \Reff. The boundaries of the boxes indicate the interquartile ranges of the data, and central lines show the median values.}
    \label{fig:morphology}
\end{figure}

The red boxes in the top panel of Fig.~\ref{fig:morphology} summarize the distribution of Sérsic index of galaxies with high kinematic misalignment within the three stellar mass bins defined in Section~\ref{SS:mstar_ssfr}. The blue boxes display the same information for aligned galaxies. This figure shows that most misaligned galaxies have a Sérsic index larger than two, hence they preferentially exhibit early-type morphologies characterised by a centrally concentrated stellar mass distribution. In contrast, the aligned sample is dominated by galaxies with lower Sérsic index, typically more associated with pseudo-bulges and late-type morphologies. The differences observed between the distributions of the misaligned and the aligned subsets across stellar mass are statistically significant as confirmed by KS tests that yield $p$-values $\ll 0.01$. For both subsets, a trend of increasing Sérsic index with stellar mass is observed. 

Likewise, the middle panel of Fig.~\ref{fig:morphology} summarizes the distributions of observed ellipticity, calculated as $1-b/a$, for misaligned (red) and aligned (blue) galaxies as a function of stellar mass. Note that the observed ellipticity is a combination of intrinsic shape and projection effects. However, since we are only interested in an internal comparison between different subsets of our sample, whose selection functions are expected to be uncorrelated with projections effects, we can use the observed ellipticity as a proxy of intrinsic shape. By working with the observed ellipticity, the KS tests comparing the distributions of misaligned and aligned galaxies across stellar mass return $p$-values of $0.67$, $0.07$, and $0.002$, respectively, for the low-, intermediate-, and high-mass bins. Nevertheless, the main purpose of this test is to show that the ellipticity of misaligned galaxies tends to decrease with stellar mass, suggesting that massive systems are more spherically symmetric compared to lower-mass galaxies.

The significance of both trends for the misaligned subset---i.e., the increase in Sérsic index and the decrease in ellipticity with stellar mass---seen in Fig.~\ref{fig:morphology} is confirmed by KS tests, which yield $p \ll 0.01$ when comparing the low‑ and high‑mass bins. Together with Fig.~\ref{fig:ssfr_vs_compactness}, this result supports the idea that high-mass misaligned galaxies have more centrally concentrated stellar mass distributions than their lower-mass counterparts. Consequently, these galaxies exert weaker stellar torques on kinematically misaligned gas, leading to longer relaxation times for unstable configurations \citep{Lake1983}.

Finally, we analyse the distributions of central stellar velocity dispersion, $\sigma_{*\; 0.25}$, as a function of stellar mass for both subsets. To do so, we estimate $\sigma_{*\; 0.25}$ for each galaxy by determining the median value of velocity dispersion from all the unmasked spaxels within a central circularized aperture of $0.25$ \Reff. The bottom panel of Fig.~\ref{fig:morphology} shows that misaligned galaxies have, on average, higher $\sigma_{*\; 0.25}$ than aligned galaxies of similar mass---in line with the idea that kinematically misaligned gas structures are more common in ETGs. These differences are supported by KS tests that yield $p$-values $\ll 0.01$ regardless of the stellar mass. As \citet{Bluck2020a} show, central velocity dispersion is a strong predictor of quenching. The higher values we find in misaligned galaxies therefore suggest that these systems have experienced more substantial quenching than their aligned counterparts. This, in turn, supports the scenario in which most misalignment events replenish the gas reservoir of galaxies through external accretion.

\bsp	
\label{lastpage}
\end{document}